\documentclass[twocolumn,showpacs,amsmath,preprintnumbers,aps,prb]{revtex4-1}
\usepackage{graphicx}  
\usepackage{dcolumn}
\usepackage{bm}

\begin{document}

\title{Coplanar stripline antenna design for optically detected magnetic resonance on semiconductor quantum dots}  

\author{F. Klotz,$^1$ H. Huebl,$^2$ D. Heiss,$^1$ K. Klein,$^1$ J. J. Finley,$^1$ and M. S. Brandt$^1$}
\affiliation{$^1$Walter Schottky Institut, Technische Universit\"{a}t M\"{u}nchen, Am Coulombwall 3, 85748 Garching, Germany}
\affiliation{$^2$Walther-Meissner-Institut, Bayerische Akademie der Wissenschaften, Walther-Meissner-Str.~8, 85748 Garching, 85748 Garching}

\begin{abstract}
\noindent We report on the development and testing of a coplanar stripline antenna that is designed for integration in a magneto-photoluminescence experiment to allow coherent control of individual electron spins confined in single self-assembled semiconductor quantum dots. We discuss the design criteria for such a structure which is multi-functional in the sense that it serves not only as microwave delivery but also as electrical top gate and shadow mask for the single quantum dot spectroscopy. We present test measurements on hydrogenated amorphous silicon, demonstrating electrically detected magnetic resonance using the in-plane component of the oscillating magnetic field created by the coplanar stripline antenna necessary due to the particular geometry of the quantum dot spectroscopy. From reference measurements using a commercial electron spin resonance setup in combination with finite element calculations simulating the field distribution in the structure, we obtain an average magnetic field of $0.2$~mT at the position where the quantum dots would be integrated into the device. The corresponding $\pi$-pulse time of $\approx 0.3$~$\mu$s fully meets the requirements set by the high sensitivity optical spin read-out scheme developed for the quantum dot.\\

\end{abstract}

\maketitle

Single electron spins confined in semiconductor nanostructures such as quantum dots (QDs) are promising candidates for future quantum computation applications.\cite{Los98, Lad10} Great improvements have been made over the last two decades to initialize,\cite{Ata06} store,\cite{Hei10, Hei08, Hei09} manipulate,\cite{Kop06, Kop08} and read out \cite{Elz04, Han05, Hei10, Hei08, Vam10} electronic degrees of freedom in quantum dots. Coherent manipulation of the electron spin over timescales shorter than that over which quantum phase coherence persists is one of the key requirements to perform logical quantum operations. The most commonly used technique to manipulate electron spins in an external magnetic field $B_0$ is electron spin resonance (ESR) that employs the oscillating magnetic field component $B_1$ of a microwave (mw) field to resonantly drive transitions between the two electron spin states split by the Zeeman energy $E_Z = g \mu_B B_0$, where $g$ is the electron $g$-factor and $\mu_B$ the Bohr magneton.
This well established method dates back to 1936 and is nowadays successfully employed in many areas of solid state physics, chemistry, bio chemistry, and materials science.\cite{Gor36, Zav45} Typically, all these applications deal with large ensembles of electron spins. Transferring ESR to the single spin level is experimentally challenging and the necessary high sensitivity is e.g.~achieved by converting the spin information to a charge information via a spin-dependent charging or transport process,\cite{Hei10, Hei08, Kop06} energy selective readout \cite{Elz04, Mor10} or optical readout.\cite{Hei10, Hei08, Vam10, Jel04} 
Until now, coherent control of single electrons in a QD via microwaves has only been demonstrated in electrostatically defined semiconductor QDs.\cite{Kop06, Kop08} Due to the low $g$-factor of 0.2~-~0.4 and the resulting low Zeeman energy as well as their charging energies such devices are typically operated at millikelvin temperatures.\cite{Kop06, Pot03, Kog04} Furthermore, they are optically inactive, limiting the possible modes of operation to electrical control of the system.

In the experiments using electrostatically defined QDs, the microwaves were supplied via an on-chip antenna formed by a terminated coplanar stripline (CPS) close to the device to maximise the oscillating magnetic field at the site of the QD.\cite{Kop06} In contrast, self-assembled QDs do not require temperatures in the mK range, but can be operated up to 70~K and are optically active.\cite{Duc04, War00} For these types of dots, only one successful experiment on a single QD involving microwaves has been reported until now.\cite{Kro08} In this case, a loop antenna placed next to the sample was used, but coherent control of the electron spin was not achieved. Only very recently, this has been realised for a single, self-assembled QD, however, using an optical method to manipulate the electron spin.\cite{Pre10} In this work, we report on the development and testing of a CPS antenna that is suitable for the coherent manipulation and readout of a single electron in a self-assembled QD as proposed by Heiss et al.\cite{Hei08} We present numerical calculations of the magnetic field distribution in such a structure deduced from finite element simulations and experimentally demonstrate the feasibility of mw manipulation of electron spins via the in-plane magnetic field component of our CPS antenna by performing electrically detected magnetic resonance (EDMR) measurements on hydrogenated amorphous silicon (a-Si:H).

To introduce the design criteria for the microwave antenna we review the micro-magnetoluminescence setup used for single spin readout in self-assembled quantum dots (see Fig.~1a) including the readout protocol.\cite{Hei10, Hei08} A schematic of the sample structure and the band diagram in growth direction are presented in Fig.~1b$\:$\&$\:$c, respectively. The QDs are embedded within the intrinsic region of a Schottky photodiode formed by a heavily \textit{n}$^+$-doped back contact and a semitransparent titanium top contact. This geometry allows the application of electric fields along the growth direction of the structure enabling tuning of the tunneling times of the photogenerated charge carriers from the dot to the corresponding electrical contacts. To facilitate optical addressing of individual QDs, the Ti layer is covered with an opaque Au film to which 1~-~5~$\mu$m wide apertures are opened. A circularly polarised laser pulse is used to resonantly excite the QD in the orbital ground state creating an exciton with predefined spin configuration. The structure is biased such that the hole tunnels out of the dot rapidly whilst the electron remains trapped (see Fig.~1c) over millisecond timescales and beyond.\cite{Hei08, Hei09} This is made possible by an AlGaAs layer underneath the QDs that forms a potential barrier for the electron leading to an asymmetric tunneling behaviour for the two charge carriers. After the tunneling of the hole, the spin system is prepared and the coherent manipulation should be performed. 
For readout a spin-to-charge conversion scheme is used mapping the spin orientation of the electron onto the charge occupancy of the dot which can then be probed in a standard micro-photoluminescence measurement.\cite{Hei10, Hei08} The experiment is performed in Faraday configuration and the sample is illuminated under normal incidence. The luminescence signal is collected via the same optical path and is spectrally and temporally separated from the excitation pulse.

\begin{figure}[h] 
	\centering
		\includegraphics[width=0.48\textwidth]{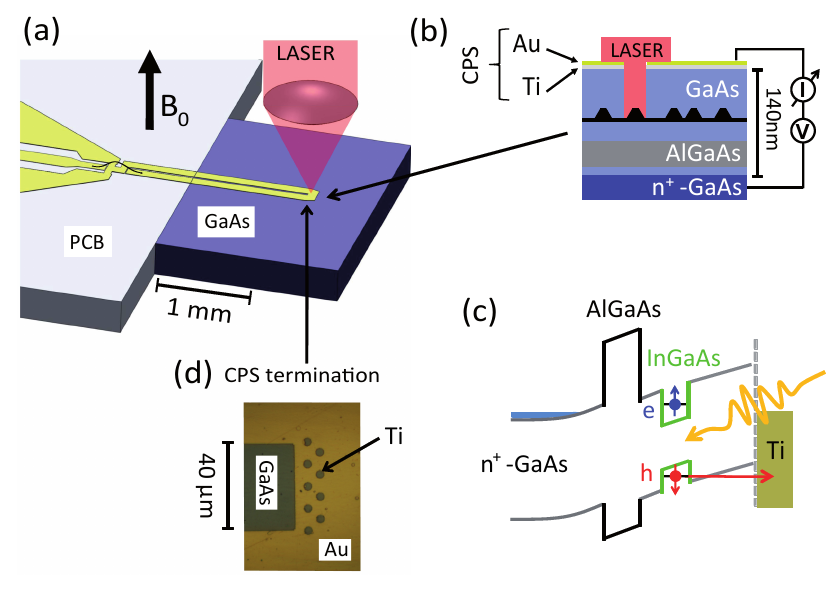}
	\label{fig:graph001}
	\caption{(colour online) (a) Illustration of the micro magneto-photoluminescence experiment the CPS antenna examined in this work is designed for. The external magnetic field $B_0$ is created by a superconducting solenoid in which the sample is placed and is perpendicular to the sample surface. (b) Schematic of the sample structure. A single layer of InGaAs QDs is embedded in the intrinsic region of a Schottky photodiode. An AlGaAs layer underneath the QDs leads to an asymmetric tunneling behaviour for electrons and holes, allowing the storage of optically excited electrons in the dot. (c) Schematic of the band structure with illustration of the electron spin initialisation method. Resonant optical excitation of an exciton is followed by a rapid tunnel escape of the hole out of the dot. (d) Optical microscope image of a CPS antenna structure for implementation in a single QD micro-photoluminescence experiment. To make single QDs optically addressable, 5~$\mu$m wide, lithographically defined apertures are opened to the gold film at the position of the desired oscillating magnetic field created by the antenna.}
\end{figure}

In our design, the CPS antenna is not only used to deliver the microwave field to the QD but also forms the electrical top contact and the shadow mask which leads to partly contradicting demands for the geometry of the antenna: (i) Since the self-assembled QDs are randomly distributed over the sample, only a small percentage of apertures are placed over exactly one single QD, while others may contain several or no dots. Therefore, a large area of the stripline termination for potential optical measurements containing many apertures is desirable to increase the chance of finding a suitable aperture. (ii) For the application of electric fields, a small area for the whole CPS structure is required to decrease the RC time constant $\tau_{RC}$, allowing fast and accurate voltage switching (with $\tau_{RC}$ ideally $< 100$~ns ). Besides a reduction of the stripline termination area, the RC time constant can be further reduced by introducing an electrically insulating layer of e.g. benzocyclobutene (BCB) or SiO$_2$ between the area of the CPS that does not contain apertures for the QD spectroscopy and the GaAs. (iii) A small area for the stripline termination is also wanted since this results in a higher current density leading to a larger amplitude of the oscillating magnetic field $B_1$ and, therefore, allowing for faster electron spin rotations. However, an increase of the field amplitude comes at the cost of a decrease in the area of the sample available for potential measurements. (iv) The magnetic $B_1$ field component excited by the CPS antenna has to be perpendicular to $B_0$ and, therefore, for this particular experimental geometry in the plane of the sample. For a shorted CPS antenna this $B_1$ field orientation is found beneath the CPS structure and particularly underneath the short as described by the Biot-Savart law. Advantageously, this allows to use the CPS structure as a shadow mask and electrical contact by positioning it directly on top of the QD. The in-plane component required of the magnetic field sets this experiment aside from other CPS antenna applications such as electron spin manipulation in electrostatically defined QDs \cite{Kop06} or electrically detected magnetic resonance of phosphorus donors in silicon \cite{Bev08,Dre11} that employ the commonly used out of-plane component of $B_1$. Figure~1d shows an optical microscope image of a first realization of the termination area of such a structure which satisfies the requirements stated above and which allowed storage of single electrons as observed in standard micro-photoluminescence experiments, albeit without the application of microwaves.\cite{Bsp1}

We start by simulating the CPS structure. The colour coded image plots in Fig.~2 show the calculated relative magnetic field distribution generated by the CPS antenna in a plane parallel to and 100~nm beneath the sample surface, the position where the optically active quantum dots are typically located. The results were obtained from finite element calculations using the software package COMSOL. For the simulation, a high frequency AC voltage with $\nu = 9$~GHz is applied to the two ports of the antenna structure. As depicted in Fig.~2, in this geometry an in-plane field is expected underneath the gold antenna structure in the vicinity of the short. Implementation of the apertures to the gold film as presented in Fig.~1c into the simulation did not result in any significant change in the field distribution and, thus, are not included in the presented figure for the sake of clarity. From the simulations we also find that the electric component of the microwave field has a node at the site of the CPS termination, minimising perturbations of the QD system by electric fields.

\begin{figure}[h]
	\centering
		\includegraphics[width=0.46\textwidth]{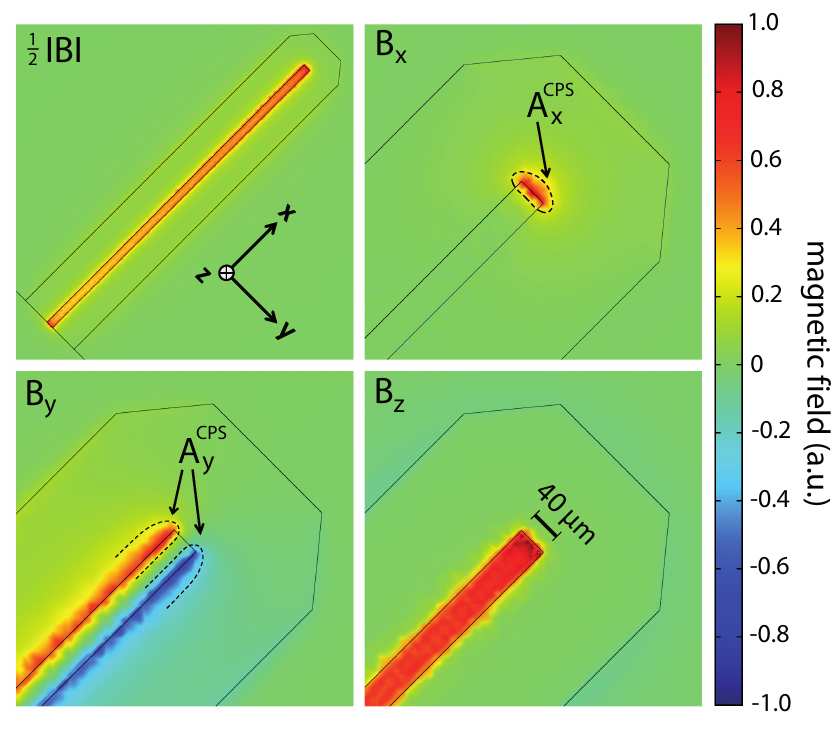}
	\label{fig:graph002}
	\caption{(colour online) Calculated distribution of the magnetic field in the plane 100 nm below the Au film forming the CPS antenna at the position where the optically active quantum dots are located. $B_{x(y)}$ is the in-plane and $B_z$ the out-of-plane component of $B$. The dashed lines enclosing the area $A_{x(y)}^{CPS}$ are marking the border at which $B_{x(y)}$ is reduced to $1/e$ of its maximum value. The field distributions are determined from finite element simulations performed with the software package COMSOL for microwave excitation of the CPS antenna structure with $\nu = $ 9 GHz. For $\left|B\right|$ in the upper left panel, the field amplitudes were divided by a factor two to allow the use of the same colour code scale for all four graphs.} 
\end{figure}

\begin{figure}[b]
	\centering
		\includegraphics[width=0.46\textwidth]{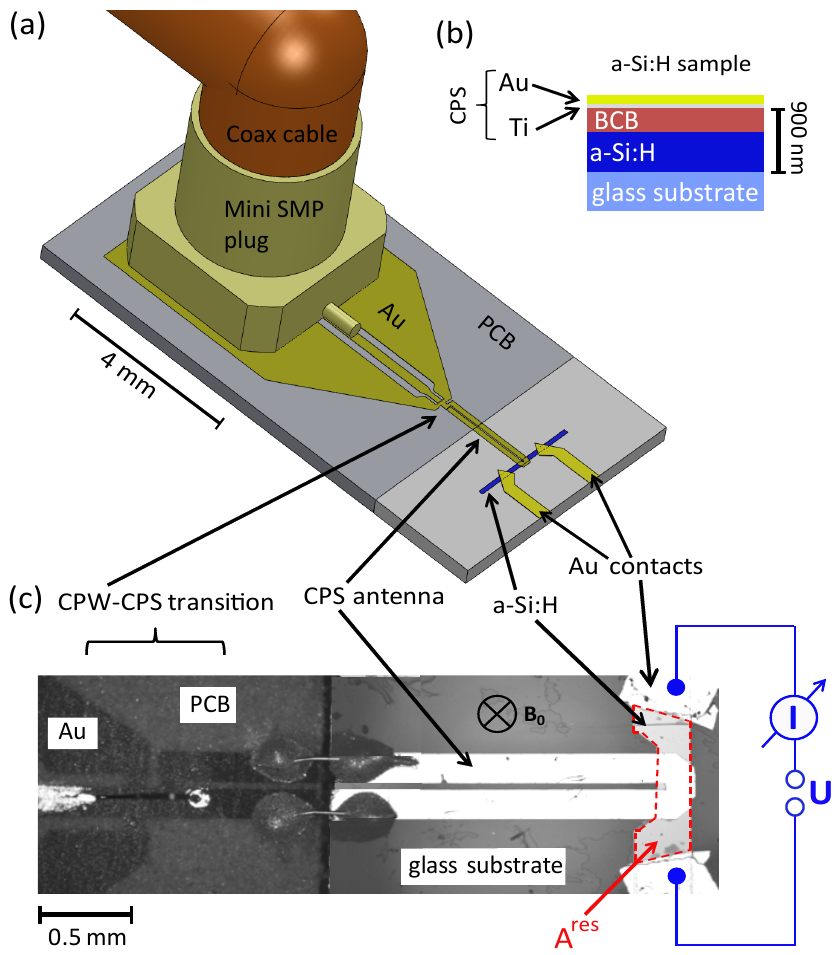}
	\label{fig:graph001}
	\caption{(colour online) (a) To-scale model of the experimental setup used in this work. (b) Schematic of the sample structure. A 400~nm thick electrically insulating benzocyclobutene (BCB) spacer is placed between the Ti/Au film forming the CPS and the hydrogenated amorphous silicon thin film (a-Si:H). (c) Optical microscope image of the experimental setup used to measure the EDMR spectrum presented here to investigate the performance of the CPS antenna. The CPS is terminated on top of the 500 nm thick a-Si:H strip allowing the creation of oscillating in-plane magnetic fields. The photocurrent signal is detected via two Au contacts on the a-Si:H film with a distance of $\approx 0.7$~mm.}
\end{figure}

To test the antenna design experimentally for spin resonance experiments where the in-plane component is essential, we use a-Si:H, a material known to exhibit a strong ESR-induced signal in the electric transport. Figure~3a shows a to-scale model of the of the experimental setup used in this work. The layer structure of the sample is presented schematically in Fig.~3b. A 500~nm thick, 1~mm long, and 0.3~mm wide strip of intrinsic hydrogenated amorphous silicon was fabricated from a 500~nm thick layer of plasma-enhanced chemical vapour deposition grown a-Si:H on a glass substrate by photolithography and reactive ion etching and subsequently covered with a 400~nm thick, electrically insulating layer of BCB photoresist. On top of this, the CPS antenna structure was defined from a 30~nm titanium adhesion layer and a 200~nm thick gold layer, that were both thermally evaporated using optical lithography. 
The lateral geometry of the sample and the microwave delivery system can be seen from the optical microscope image presented in Fig.~3c. The short of the striplines that defines the region for electron spin resonance measurements was placed directly over the center of the a-Si:H strip that is electrically contacted via two lithographically defined Au contacts with a distance of $\approx$ 0.7~mm. The microwaves are delivered to the sample via a CPS formed by a 400~nm thick Au film sputtered on a printed circuit board (PCB, Rogers RO3010) where the Cu layer has been fully removed in this area by H$_2$O$_2$/HCl etching. The microwaves are fed into the CPS on the PCB via a coplanar waveguide to stripline (CPW-CPS) transition \cite{Kim02, Ana08, But04} which is a slightly modified and scaled-up version of the one presented in Ref.~
25. The CPW itself is contacted via a standard PCB mount-on microwave plug (Rosenberger Mini SMP 18S102-40ML5) which is soldered to a low loss semirigid coax cable (see Fig.~3a). The connection between the two CPSs on the PCB and the sample is formed by Au wires that were glued to the CPSs on both sides with conducting Ag epoxy resin. The whole sample structure was placed in a magnetic field of $B_0 \approx 320$~mT applied perpendicular to the sample surface as in the QD experiment. Illumination of the a-Si:H was achieved from the backside through the glass substrate using the white light of a halogen lamp.

The detection principle of EDMR is based on spin-dependent transport processes in the investigated material.\cite{Kap78} In a-Si:H, a reduction of the photocurrent $I$ by $\Delta I$ is observed when flips of the electron spin e.g. of paramagnetic defects are induced by a microwave $B_1$ field.\cite{Sol77} Figure~4a presents the result of such a measurement at room temperature. Microwaves at a frequency of $\nu = 9$~GHz were applied via the CPS antenna to the a-Si:H film using amplitude modulation to employ lock-in detection. A clear resonance is observed at 320.61~mT. The resonance position corresponds to a $g$-factor of $2.0056 \pm 0.0001$ as expected for dangling bonds in a-Si:H, the dominant signal detected in EDMR of intrinsic a-Si:H thin films with low defect density at room temperature.\cite{Str82, Der83}
The lineshape and linewidth of the dangling bond resonance are governed mostly by the anisotropy of the $g$-factor of dangling bonds as well as the random orientation of these bonds in the amorphous material leading to a so-called powder pattern.\cite{Stu89, Feh} Since we are here only interested in the EDMR signal intensity, we use a simple Lorentzian to describe the resonance. A fit of this lineshape to the spectrum in Fig.~4a shows that the relative change in the photocurrent monitored is $\delta^{CPS}=- \Delta I / I \approx 1.6 \cdot 10^{-5}$ and, after averaging over six measurements, exceeds the noise level  by one order of magnitude.

\begin{figure}[h]
	\centering
		\includegraphics[width=0.4\textwidth]{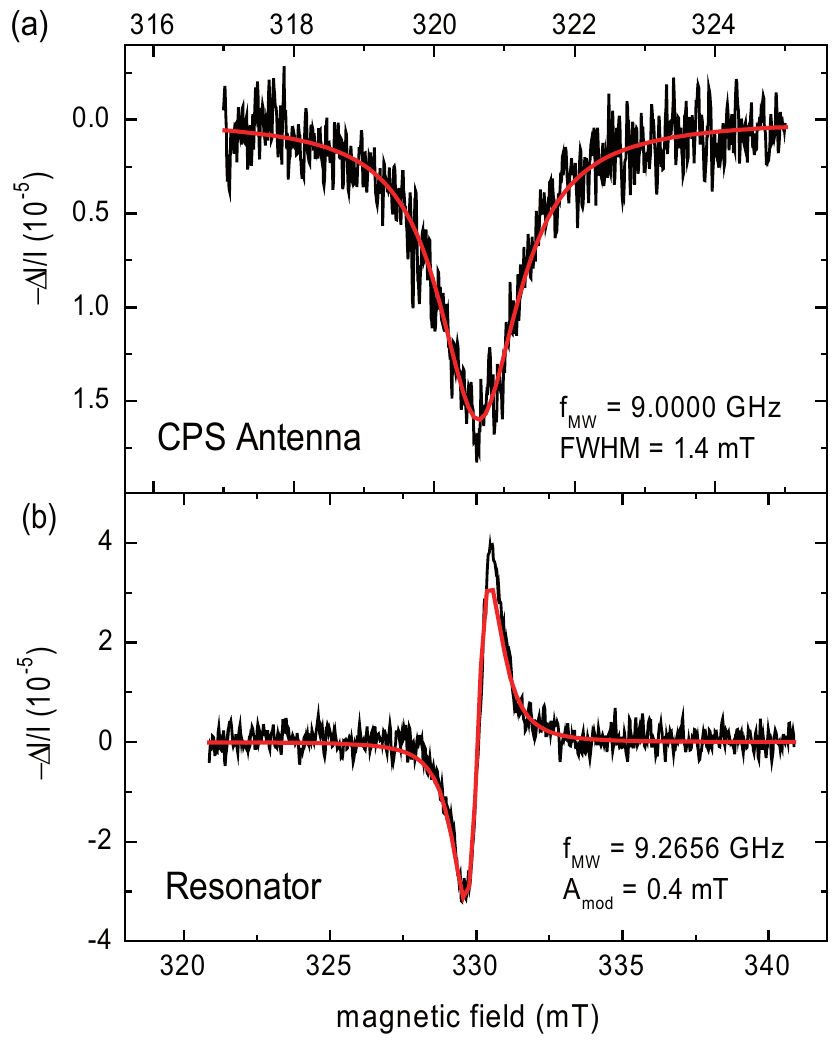}
	\label{fig:graph003}
	\caption{(colour online) (a) EDMR signal of a-Si:H measured with the CPS antenna structure shown in Fig.~3c using mw-amplitude modulation for lock-in detection. (b) EDMR signal of the same sample using the same electrical contacts for photocurrent detection measured in a commercial TE$_{\rm{102}}$ microwave resonator employing magnetic field modulation for lock-in detection.}
\end{figure}

To obtain a quantitative measure for the magnetic field generated by the CPS antenna $B_1^{CPS}$, we compare it to the $B_1$ field generated in a $\lambda-$cavity $B_1^{res}$ which is more conventionally used for EDMR. For this measurement, the glass sample with the a-Si:H film and the electrical contacts was removed from the PCB sample holder and placed into a commercial TE$_{\rm{102}}$ resonator where the $B_1$ field strength  is known. The corresponding EDMR spectrum taken is shown in the Fig.~4b. Since for this experiment magnetic field modulation in combination with lock-in detection was used, the obtained signal is the derivative of the Lorentzian lineshape. 
Correcting for the modulation amplitude of 0.4~mT used for a FWHM linewidth of 1.4~mT, corresponding to a peak-to-peak linewidth of 0.8~mT (see Fig.~4a$\:$\&$\:$b), \cite{Poole} the detected signal has a signal amplitude of $\delta^{res} = -\Delta I / I \approx 17 \cdot 10^{-5}$ which is about one order of magnitude larger than the signal obtained with the CPS antenna. 
The observed signal-to-noise ratio is of similar size for both methods.

The applied microwave power in the experiment using the CPS antenna was $P^{CPS} = 300$~mW. However, test measurements on the microwave delivery lines on the PCB showed an attenuation of $\approx 4$~dB at $\nu=9$~GHz, resulting in only $\approx 120$~mW delivered to the CPS antenna. The microwave power used for the reference experiment was $P^{res} = 200$~mW generating a $B_1$ field amplitude of $B_1^{res}=0.066$~mT at the site of the sample in the TE$_{\rm{102}}$ resonator,\cite{Stu86, Kaw97} a regime in which the amplitude of the EDMR signal is proportional to the $B_1$ field.\cite{Kaw97} 
While for the EDMR experiment in the resonator the whole a-Si:H film in between the Au contacts with an area $A^{res}$ (indicated by the red dashed line in Fig.~3c) contributes to the signal, only a small fraction $A^{CPS}$ (indicated by the black dashed lines in Fig.~2) of the material is exposed to a sufficiently large mw field when the CPS antenna is used as can be seen from Fig.~2. To estimate this geometric factor, we compare the area $A^{CPS}$ for which $B_1^{CPS}$ is above $1/e$ of its maximum value determined from the calculated magnetic field distribution presented in Fig.~2 with the whole area $A^{res}$ of a-Si:H in between the electrical contacts.\cite{info1} Only the field components $B_x$ and $B_y$ have to be taken into account for this purpose as the sample is positioned such that $B_z$ is parallel to $B_0$ and, therefore, does not contribute to the EDMR signal. Using this approach, we find a ratio of $A^{CPS} / A^{res}\approx 1 / 20$. Taking all these factors into account, we finally obtain $B_1^{CPS} =  \delta^{res} / \delta^{CPS} \cdot P^{res}/P^{CPS} \cdot A^{res} / A^{CPS} \cdot B_1^{res} \approx 0.2$~mT for a microwave power of $\approx 120$~mW at the beginning of the mw delivery line.

To estimate the applicability of this approach to coherent manipulation of electron spins in self-assembled quantum dots via pulsed ESR, we have to compare the spin-lattice relaxation time $T_1$ of an electron in a quantum dot with the $\pi$-pulse time $t_{\pi}$ achievable. For a $B_1$ field of 0.2~mT, we obtain $t_{\pi} = h / (2 g \mu_B B_1) \approx 0.3$~$\mu$s using $\left|g\right| =$ 0.6 as measured for typical self-assembled InGaAs quantum dots.\cite{Kro08} In these dots, $T_1$ depends strongly on the applied magnetic field $B_0$ but ranges between 600 to 50~$\mu$s for electron and 400 to 5~$\mu$s for hole spins for magnetic fields of 5~T~$<B_0<$~12~T at a temperature $T = 8$~K, a typical temperature for the spectroscopy of single self-assembled QDs.\cite{Hei07} However, since $T_1$ is proportional to $T^{-1}$, even longer $T_1$ times are possible by further reducing the temperature.\cite{Hei07, Kro04} At 1~K and 4~T, electron spin relaxation times as long as 20~ms have been demonstrated.\cite{Kro04}  
Another important timescale is the storage time $t_s$, describing the time the electron can be stored in the QD. Tunneling escape of the electron out of the dot strongly depends on the electric field applied to the sample and can be tuned to timescales as long as seconds during the spin manipulation phase,\cite{Hei09} thus, not setting any limitation to the measurement since $T_1 < t_s$ is always achievable.
A comparison clearly shows that $t_{\pi}$ is significantly shorter than $T_1$ and that our approach should be very suitable for spin control of single electrons confined in self-assembled QDs even taking into account large errors margins for the estimation of $B_1^{CPS}$ above.

For the frequency of $\nu = 9$~GHz where the operation of the CPS antenna is demonstrated, the corresponding resonance field is $B_0 \approx 1$~T for which even longer $T_1$ times in the range of seconds for the electron in the QD are expected.\cite{Kro04} However, the CPS antenna is a purely non-resonant structure and, therefore, should allow broadband operation. Test measurements showed that the whole microwave delivery system including the CPW to CPS transition and the mount-on devices on the PCB allows operation up to 35~GHz with losses of less than 10~dB offering the possibility to conduct ESR at $B_0$ fields of 3.5~T where $T_1$ is still three orders of magnitude longer than $t_{\pi}$ at 8~K.\cite{Hei07} This should allow to investigate the magnetic field dependence of electron spin dynamics via coherent spin manipulation with the CPS antenna over a wide range of several Tesla and could provide valuable information for the identification of decoherence mechanisms in quantum dots.
Performing ESR experiments on self-assembled QDs at higher $B_0$ fields also has practical benefits since a pair of spectrally resolvable Zeeman states allows to address the two spin configurations via their energy, removing the need for perfectly circularly polarised light for spin initialisation and spin-to-charge conversion.\cite{Hei10}
Furthermore, the fidelity of the spin initialisation, which is achieved by resonant optical excitation of an exciton and subsequent removal of the hole from the dot, potentially increases with increasing $B_0$. This is due to the anisotropic exchange interaction that excitons experience in self-assembled QDs with imperfect symmetry which leads to a fine-structure splitting of the exciton states at $B_0=0$~T, typically in the range of $\sim$ 30~-~150~$\mu$eV for self assembled InGaAs QDs.\cite{Bsp1, Bay02} If the exchange coupling is larger than the Zeeman energy and the extraction of the hole does not occur over sufficiently fast timescales, the eigenstates are of strongly mixed spin character in which case spin initialisation with a high degree of polarization is not achievable.

In conclusion, we demonstrated the feasibility of electron spin manipulation using the in-plane component of the oscillating magnetic field created by a coplanar stripline antenna. Furthermore, the structure is designed such that it can simultaneously serve as an electrical top contact and opaque shadow mask in a single self-assembled quantum dot spectroscopy experiment. From EDMR reference measurements on amorphous silicon in combination with finite element simulations of the field distribution in the structure, we obtain an average in-plane $B_1$ field of $\approx 0.2$~mT at the site of the dots. The corresponding $\pi$-pulse time $\approx 0.3$~$\mu$s for self-assembled InGaAs QDs with $\left|g\right|=$~0.6 is several orders of magnitude shorter than required and should, therefore, make spin manipulation in single self-assembled QDs possible. 

We would like to thank F. Finger and O. Astakhov at the Forschungszentrum J\"ulich for the growth of the a-Si:H films.  
This work was financially supported by the DFG via SFB 631, Project C6, the work in J\"ulich by the BMBF via EPR-Solar. \\


\begin{thebibliography}{Abc99}
\bibitem{info1} The calcualtions show that this is possible since there is no significant difference in relative and absoulte field distribution between 100~nm and 1~$\mu$m beneath the sample surface.
\end{thebibliography}
\end{document}